\documentclass[fleqn,usenatbib]{mnras}

\usepackage[T1]{fontenc}

\DeclareRobustCommand{\VAN}[3]{#2}
\let\VANthebibliography\thebibliography
\def\thebibliography{\DeclareRobustCommand{\VAN}[3]{##3}\VANthebibliography}

\usepackage{graphicx}	% Including figure files
\usepackage{amsmath}	% Advanced maths commands
\usepackage{gensymb}
\usepackage[dvipsnames]{xcolor}

\usepackage{newtxtext,newtxmath}
\usepackage[normalem]{ulem}
\title[Accretion from a polar circumbinary disc]{Accretion onto a binary from a polar circumbinary disc}

\author[Smallwood et al.]{Jeremy L. Smallwood,$^{1,2}$\thanks{E-mail: drjeremysmallwood@gmail.com}
Stephen H. Lubow,$^{3}$ 
and Rebecca G. Martin$^{2}$
\\
$^1$Center for Astrophysics, Space Physics, and Engineering Research, Baylor University, Waco, Texas 76798-7316, USA\\
$^2$Department of Physics and Astronomy, University of Nevada, Las Vegas, 4505 South Maryland Parkway, Las Vegas, NV 89154, USA\\
$^3$Space Telescope Science Institute, Baltimore, MD 21218, USA
}

\date{Accepted XXX. Received YYY; in original form ZZZ}

\pubyear{2020}

\begin{document}
\label{firstpage}
\pagerange{\pageref{firstpage}--\pageref{lastpage}}
\maketitle

\begin{abstract}
We present hydrodynamical simulations to model the accretion flow from a polar circumbinary disc onto a high eccentricity ($e=0.78$) binary star system with near unity   mass ratio ($q=0.83$), as a model for binary HD 98800 BaBb.  
We compare the polar circumbinary disc 
 accretion flow with the previously studied coplanar case.  In the coplanar case, the circumbinary disc becomes eccentric
 and the accretion alternates from being dominant 
 onto one binary member to the other.
For the polar disc case involving a highly eccentric binary, we find that the circumbinary
disc retains its initially low eccentricity and that the primary star accretion rate is always about the same as the secondary star accretion rate.   Recent observations of the binary HD 98800 BaBb, which has a polar circumbinary disc, have been used to determine the value of the $\rm H\alpha$ flux from the brighter component. From this value, we infer that the accretion rate is much lower than for typical T Tauri stars.   The eccentric orbit of the outer companion HD 98800~A increases the accretion rate onto HD 98800 B by $\sim 20$ per cent after each periastron passage.  Our hydrodynamical simulations are unable to explain such a low accretion rate unless the disc viscosity parameter is very small, $\alpha < 10^{-5}$.  Additional observations of this system would be useful to check on this low accretion rate. 
\end{abstract}


%%%%%%%%%%%%%%%%%%%%%%%%%%%%%%%%%%%%%%%%%%%%%%%%%%

%%%%%%%%%%%%%%%%% BODY OF PAPER %%%%%%%%%%%%%%%%%%

\section{Introduction}
\label{intro}

It is estimated that more than $40$-$50\%$ of stars in the galaxy are in a binary pair \citep{Duquennoy1991,Raghavan2010,Tokovinin2014a,Tokovinin2014b}. Circumbinary discs of gas and dust are common around young binary stars and are locations for planet formation.
Misalignments between the circumbinary disc and the binary orbital plane are commonly observed \citep{Czekala2019}. The pre-main-sequence binary KH~15D has a circumbinary disc tilted by about $3\degree-15\degree$  \citep{Chiang2004,Smallwood2019,Poon2020}. GG Tau A, another pre-main sequence binary, has a circumbinary disc that is inclined by about $37\degree$ \citep{Keppler2020}.  The pre-main sequence circumtriple disc around the hierarchical triple star system, GW Ori, is misaligned by about $38\degree$ \citep{Bi2020,Kraus2020,Smallwood2021b}.
The young binary IRS 43 has a circumbinary disc with an observed misalignment of about $ 60\degree$ \citep{Brinch2016}. The  $6$--$10\, \rm Gyr$ old binary system, 99 Herculis, has a nearly polar (about $87\degree$) debris ring \citep{Kennedy2012,Smallwood2020a}.

A misaligned young system of particular interest is HD 98800.
The system is at a distance of $47\, \rm pc$ \citep{VanLeeuwen2007}, and comprises two pairs of binaries,  HD 98800 AaAb  and HD 98800 BaBb. The two sets of binaries are orbiting each other with semi-major axis $54\, \rm au$, eccentricity $e_{\rm AB} = 0.52\pm 0.01$ and orbital period of $246\pm 5$ years \citep{Kennedy2019}. The orbit of the BaBb binary is well constrained with a semi-major axis $a \simeq 1\, \rm au$ and eccentricity $e = 0.785 \pm 0.005$. The masses of the two B-binary stars  are $M_{\rm Ba} = 0.699\, \rm M_{\odot}$ and  $M_{\rm Bb} = 0.582\, \rm M_{\odot}$ \citep{Boden2005}. The circumbinary disc was first thought to be coplanar to the binary orbital plane \citep{Andrews2010} but the later observations showed the disc to be misaligned. 
From their ALMA observations,  \cite{Kennedy2019} modelled the dust and carbon monoxide measurements and inferred that the radial dust component extends from $2.5 \pm 0.02\, \rm au$ to $4.6 \pm 0.01\, \rm au$ and the gas component extends from $1.6\pm 0.3\, \rm au$ to $6.4 \pm 0.5\, \rm au$. Both the dust and gas components were found to be in a similar orientation. From these observations, the circumbinary disc has an inferred inclination to the binary orbital plane of either $48\degree$ or $90\degree$ (polar). The polar configuration is more likely since their disc modelling suggests that the alignment time of a $48\degree$ disc to reach polar is very short compared to the stars' age.   Additional support of the polar configuration was provided by
the  small observed size of the inner cavity in the circumbinary disc that requires a nearly polar disc orientation \citep{Franchini2019b}.

 A misaligned circumbinary disc undergoes nodal precession. For a low initial inclination, the precession is around the binary angular momentum vector \citep[e.g.,][]{Larwoodetal1996} while for sufficiently high initial inclination, the precession is around the eccentricity vector.  Dissipation causes the disc to evolve to either align coplanar to the binary orbital plane \citep{Papaloizou1995,Lubow2000,Nixonetal2011a,Facchinietal2013,Foucart2014} or align perpendicular (i.e., polar) to the binary orbital plane \citep{Aly2015, Martinlubow2017,Lubow2018,Zanazzi2018,Martin2018}.

The dynamics of a coplanar binary-circumbinary disc system has been the subject of many studies.
The binary opens a central cavity in the circumbinary disc through the effects of its tidal torques   \citep{Artymowicz1994, Miranda2015}.
Circumbinary gas flows into the cavity in the form of gas streams \citep[e.g.,][]{Artymowicz1996,  Gunther2002,Shi2012,DOrazio2013,Farris2014, Miranda2017, Munoz2019,Mosta2019}. This flow is important in forming and/or replenishing circumstellar discs around each binary component. 
The properties of the circumbinary gas disc and flow in the central cavity vary with binary eccentricity and mass ratio.
In this paper, we are concerned with the case of a high binary eccentricity and near unity mass ratio, as occurs for HD 98800 BaBb.
%Moreover, the binary torque is related to the level of misalignment of the circumbinary disc and the eccentricity of the binary \citep{Artymowicz1994,Lubow2015,Miranda2015}. The strongest binary torque at a given radius is produced when the disc is coplanar to the binary orbital plane. The tidal torque is zero when the circumbinary disc is polar and the binary eccentricity approaches $e_{\rm b} = 1$ \citep{Lubow2018} or if the disc  is retrograde and the binary orbit is circular \cite[e.g.,][]{Nixon2013, Nixon2015}. 
%The general accretion rate through the circumbinary disc increases with the disc aspect ratio,
%as it does for typical accretion discs \citep{Pringle1981}. 
For a high eccentricity binary with mass ratio close to unity in a coplanar system, previous simulations have shown that there are
large modulations in the accretion rate towards an eccentric orbit  binary occur on the timescales of  $1 P_{\rm orb}$ (where $P_{\rm orb}$ is the binary orbital period),
due to time varying gravitational forces from the eccentric orbit binary. 
In addition, the circumbinary disc acquires substantial eccentricity around such a binary
\citep[e.g.,][]{Lubow2000a,  Munoz2016, Munoz2019}.  
As a consequence of the apsidal precession of the eccentric disc around an eccentric orbit binary with mass ratio of unity, the binary
member that accretes the most material alternates in time between the two stars \citep{Munoz2016}.

Less is known about the accretion process in the noncoplanar case. 
At a given radius, tidal torques are weaker in the noncoplanar case, giving rise to smaller central cavities \citep{Lubow2015,Miranda2015, Franchini2019b}.
Tidal torques due to Lindblad resonances on a polar disc approach zero in the limit of unity binary eccentricity \citep{Lubow2018}.
Recently, \cite{Smallwood2021a} showed that noncoplanar gas streams develop in the central cavity of a moderately inclined ($60^\circ$) circumbinary disc.
In this paper, we explore the properties of a polar circumbinary disc and accretion onto the binary as a model for  HD 98800 BaBb.

Recently, \cite{Zurlo2020} conducted a survey of 11 transitional discs with the SPHERE instrument \citep{Beuzit2019} at the Very Large Telescope (VLT). Their purpose was to undertake the largest $\rm H\alpha$ survey dedicated to protoplanets in hopes of detecting circumplanetary discs. The observations were carried out using the $\rm H\alpha$ filter of the Zimpol rapid-switching imaging polarimeter \cite[ZIMPOL;][]{Schmid2018}. The ZIMPOL is a subsystem in visible light which takes simultaneous images in the $\rm H\alpha$ narrow band filter and $\rm H\alpha$ continuum. One of the 11 systems that were observed was HD 98800 BaBb. No sub-stellar companions were detected around the binary; however, unexpectedly, an $\rm H \alpha$ flux was detected from the primary binary component Ba. No  $\rm H \alpha$ flux was detected from the secondary. From the $\rm H \alpha$ flux, \cite{Zurlo2020} estimated a $\rm H \alpha$ luminosity and an accretion luminosity. 
%The detection of an accretion luminosity can be inferred as a mass accretion rate, but \cite{Zurlo2020} does not compute this. 

In Section~\ref{sec::Hydro_Sims}, we discuss the setup of the hydrodynamical simulations. In Section~\ref{sec::disc_evolution}, we compare the disc structure of the coplanar and polar discs, along with comparing the accretion rates onto the binary.  In Section~\ref{sec::AB}, we include the outer companion HD 98800 A companion  in our hydrodynamical simulations.  In Section~\ref{sec::Mass_accretion}, we calculate the mass accretion rate onto the primary component in HD 98800 BaBb from the polar circumbinary disc using the $\rm H\alpha$ emission observations from \cite{Zurlo2020}. Finally, we present our concluding remarks in Section~\ref{sec::conclusions}.

%The most prominent example of circumbinary material flowing feeding circumstellar disc is the system [BHB2007] 11B \citep{Alves2019}.  The system has long filaments connecting the circumbinary disc to the two circumstellar discs around each binary component. The formation of accretion filaments occurs in misaligned circumbinary discs and eccentric binaries \citep{Matsumoto2019,Mosta2019}. \cite{Alves2019} estimated the mass accretion rate from the circumbinary disc onto the circumstellar discs by assuming that the material at the far end side of the filament will fall freely into the star in a straight path. Using the two streams accreting into the star, these values provide a mass accretion rate of $\sim 1.1 \times 10^{-5}\, \rm M_{\odot}/yr$. While the orientation of the [BHB2007] 11B circumbinary disc is uncertain, the relatively high accretion rate and formation of circumstellar discs suggest that this disc is in a non-polar state. There is no evidence of circumstellar discs around either binary components in HD 98800 \cite[e.g.,][]{Kennedy2019}.  The low mass accretion rate we estimate in HD 98800 gives strong evidence that the circumbinary disc is in a polar orientation. 

%\RGM{Add a paragraph on accretion rates observed in single stars/coplanar binary systems for comparison to make the point that this is very small.}

\section{Hydrodynamical Simulation Setup}
\label{sec::Hydro_Sims}

\begin{table*}
	\centering
	\caption{ Overview of the hydrodynamical simulations. The first column shows the simulation name. The second column denotes the initial disc orientation, either coplanar or polar and the third column represents the accretion radius for each binary component. The  fourth and fifth columns show the time-averaged mass accretion rate over the last $400\, \rm P_{orb}$ onto the primary and secondary stars, respectively.  The last column denotes whether the HD 98800 A companion is included.}
%	\label{tab:example_table}
	\begin{tabular}{cccccc} % four columns, alignment for each
		\hline
		& & & \multicolumn{2}{c}{$\dot{M}_{\rm acc}^{\rm avg}/(M_{\odot}/\rm yr)$} &  \\
		Simulation & Orientation & $R_{\rm acc}/\rm au$ & Primary & Secondary & Third Companion? \\
		\hline
        \hline
        run1 & coplanar & $0.25$ & $1.15\times 10^{-10}$  & $1.18\times 10^{-10}$ &  No \\
        run2 & coplanar & $0.025$ & $1.08\times 10^{-10}$  & $9.44\times 10^{-11}$ &  No \\
        run3 & coplanar & $0.01$ & $1.06\times 10^{-10}$  & $1.06\times 10^{-10}$ &  No \\
        run4 & polar & $0.25$ & $6.05\times 10^{-11}$ & $4.90\times 10^{-11}$ &  No \\
        run5 & polar & $0.025$ & $4.58\times 10^{-11}$ & $4.22\times 10^{-11}$ &  No  \\
        run6 & polar & $0.01$ & $4.43\times 10^{-11}$ & $4.35\times 10^{-11}$ &  No  \\
          run7 &   polar &   $0.25$  &  $ 6.28\times 10^{-11}$  & $ 5.31\times 10^{-11}$ &  Yes \\
        \hline
	\end{tabular}
    \label{table1}
\end{table*}

This section describes the 3D hydrodynamical simulation setup of a circumbinary disc around the eccentric orbit binary, HD 98800 BaBb.
 To model the accretion rate onto each binary component, we use the smoothed particle hydrodynamical code {\sc Phantom} \citep{Lodato2010,Price2010,Price2018}.  We consider two different initial circumbinary disc orientations, coplanar and polar.  {\sc Phantom} has been successful in  modeling misaligned circumbinary discs \citep[e.g.,][]{Nixon2012,Nixon2013,Dougan2015,Facchini2018, Aly2020,Smallwood2020a}. The hydrodynamic discs that we model here are in the wave-like regime, meaning that the disc aspect ratio, $H/R$, is larger than the \cite{Shakura1973} $\alpha$-viscosity coefficient. This is appropriate for protoplanetary discs. The consequence of this is that the warp induced in the disc by the binary torque propagates as a pressure wave with speed $c_{\rm s}/2$ \citep{Papaloizou1983,Papaloizou1995}, where $c_{\rm s}$ is the sound speed.  

The binary is modelled with the mass and orbital parameters from \cite{Kennedy2019} that are given in Section \ref{intro}.
% The primary mass is  $M_1 = 0.699\, \rm M_{\odot}$ and the secondary mass is $M_2 = 0.582\, \rm M_{\odot}$.  The binary orbit has semi-major axis $a_{\rm b} = 1\, \rm au$ and the eccentricity is  $e_{\rm b} = 0.785$. 
 When a particle enters the accretion radius, it is considered accreted and the particle mass and angular momentum are added to the sink particle.  For both the initially coplanar and the initially polar models, we consider accretion radii $R_{\rm acc} = 0.25,\, 0.025,\, 0.01\, \rm au$.  We examine various accretion radii of each binary component. The smallest accretion radius used is comparable to the size of the star.  Table~\ref{table1} gives a summary of all the hydrodynamical simulations.

The discs initially consist of $1\times 10^6$ equal mass smoothed particle hydrodynamic (SPH) Lagrangian particles that are radially distributed from the inner disc radius $R_{\rm in}$ to the outer disc radius $R_{\rm out}$. Observations show that the inner radius of the polar disc in HD 98800 BaBb is about $1.6\, \rm au$ \citep{Kennedy2019}, whereas a coplanar disc around the same binary would have a much larger inner truncation radius of about $3\,\rm au$ \citep{Artymowicz1994}.  As discussed in Section~\ref{intro}, a misaligned circumbinary disc can radially extend closer to the binary due to the binary torque being weaker at a given radius. %\cite[e.g.,][]{Lubow2015,Miranda2015,Nixon2015,Lubow2018,Franchini2019b}. 
We initially set the inner radius to be $R_{\rm in} = 4\, \rm au$ in all of our simulations since this is larger than the tidal truncation radius in all cases. This allows the material to initially move inwards before the disc reaches a quasi-steady state. If the disc were to begin too close to the binary, there could be an initially artificially enhanced accretion rate onto the binary, resulting in lost disc mass. 

The outer disc radius is set to $R_{\rm out} = 6.4\, \rm au$,    which is motivated by the observations \citep{Kennedy2019}.   We simulate the observed gas disc mass of $M_{\rm d} \sim 8\times 10^{-7}\, \rm M_{\odot}$ for the system based on CO flux  \cite[e.g.,][]{Kennedy2019}\footnote{The simulations are initially modelled with a mass of $M_{\rm d} = 10^{-3}\, \rm M_{\odot}$, but we re-scale the masses in the simulations by a factor of $8\times 10^{-4}$. We expect that this procedure is valid, since the disc mass should not play an essential role in the dynamics of the binary.}.
 The mass of the disc must be small, since the disc is observed to be only a few degrees away from a $90^\circ$ polar state, and a larger disc mass leads to a generalised polar state at lower levels of misalignment \citep{Martin2019}. %We ignore the effect of self-gravity since it does not affect the nodal precession rate of a flat circumbinary disc.
This inferred disc mass limit for HD 98800 BaBb is not large enough for self-gravity to be important. 
%{\bf \cite{Kennedy2019} derived a gas mass for the system based on CO flux of $\sim 8\times 10^{-7}\, \rm M_{\odot}$; therefore, we re-scale our densities in the simulations by a factor of $8\times 10^{-4}$. This procedure is only valid when the disc mass does not play an essential role in the dynamics.}

% \begin{figure} 
%\centering
%\includegraphics[width=1\columnwidth]{mass.eps}
%\centering
%\caption{The disc mass as a function of time for a coplanar disc (black run1 from Table ~\ref{table1}), polar disc (blue, run4), $i_0 = 10\degree$ disc (red, run7), and $i_0 = 60\degree$ disc (yellow, run8). The polar orientated discs maintain more mass versus a coplanar/low inclination disc.}
%\label{fig::mass}
%\end{figure}

The disc viscosity is modeled by using the artificial viscosity $\alpha^{\rm av}$, which is implemented in {\sc Phantom} \citep{Lodato2010}. The disc surface density profile is initially a power-law distribution $\Sigma \propto R^{-3/2}$. We adopt the locally isothermal equation of state of \cite{Farris2014} and set the sound speed $c_s$ to be
\begin{equation}
    c_{\rm s} = c_{\rm s0}\bigg( \frac{a_{\rm b}}{M_1 + M_2} \bigg)^q \bigg( \frac{M_1}{R_1} + \frac{M_2}{R_2}\bigg)^q,
    \label{eq::EOS}
\end{equation}
where $R_{\rm 1}$ and  $R_{\rm 2}$ are the radial distances from the primary and secondary star, respectively, and $c_{\rm s0}$ is a constant determined by the disc aspect ratio scaling.
%ALREADY defined above: The binary has a separation of $a_{\rm b}$ and a total mass of $M = M_1 + M_2$, where $M_1$ and $M_2$ are the mass of the primary and secondary star, respectively. 
This prescription for the sound speed distribution ensures that the primary and secondary star irradiation sets the temperatures around the circumprimary and circumsecondary disc, respectively. %\citep{Farris2014,Franchini2019a,Smallwood2021}. 
For $R_1(R_2) \gg a$, the sound speed is set by the distance from the binary centre of mass.  The disc aspect ratio $H/R$ is set to be $0.1$ at $R_{\rm in}$. With this prescription, the shell-averaged smoothing length per scale height $\langle h \rangle / H$ and the disc viscosity parameter $\alpha$ are constant over the radial extent of the disc \citep{Lodato2007}. We take the \cite{Shakura1973} $\alpha_{\rm SS}$ parameter to be $0.01$. The circumbinary disc is initially resolved with average smoothing length $\langle h \rangle / H = 0.19$.

%Material from the circumbinary disc flows through the binary cavity and promotes the formation of circumstellar discs around the binary components. 
%The streams connecting the circumbinary disc to the circumstellar disc continuously feed the circumstellar discs with material. The formation of accretion streams also occurs in misaligned circumbinary discs with eccentric binaries \citep[e.g.,][]{Smallwood2021}. 
To model the circumstellar discs in {\sc phantom} that are fed by circumbinary gas, we would need to adopt the prescription detailed in \cite{Smallwood2021a} in which we artificially reduce the sound speed close to each binary component. This increases the viscous timescale in the circumstellar discs and allows material to build up around the binary components.  However, with this prescription we would not be able to simulate for the long timescale required to model the quasi-steady state accretion rate from the circumbinary disc to the binary. Resolving bound material on short length-scales is quite computationally expensive. Therefore, we do not try to capture the formation of such discs in these hydrodynamical simulations, since the larger number of particles required to resolve the formation of circumstellar discs  around each binary component would significantly increase the computational time. The circumstellar discs may buffer accretion onto the stars, but this effect is not included since we are not revolving the individual discs.   There is no observational evidence of circumstellar discs around either of the binary components in HD 98800 BaBb \cite[e.g.,][]{Kennedy2019}, and therefore we focus on how much material is accreted onto the binary components.

\section{Coplanar and Polar Disc Evolution}
\label{sec::disc_evolution}

% \begin{figure*} 
% \centering
% \includegraphics[width=0.69\columnwidth]{theta_250.eps}
% \includegraphics[width=0.69\columnwidth]{theta_300.eps}
% \includegraphics[width=0.69\columnwidth]{theta_350.eps}
% \includegraphics[width=0.69\columnwidth]{theta_400.eps}
% \includegraphics[width=0.69\columnwidth]{theta_450.eps}
% \includegraphics[width=0.69\columnwidth]{theta_500.eps}
% \centering
% \caption{We show the azimuthal surface density of a coplanar circumbinary disc at six different evolutionary times beginning with $250\, \rm P_{orb}$ and progressing in steps of $50\, \rm P_{orb}$. The $x$--axis denotes the disc radius, $r$, while the $y$--axis represents the azimuthal angle, $\theta$. The colorbar shows the surface density, with red being the most dense and white being the least. The over-dense clump is located with a separation of $\sim 5\, \rm au$ and  orbits around the binary stars. \RGM{This figure is not necessary. I don't think it adds any new information to figure 4? }  } 
% \label{fig::theta}
% \end{figure*}

\begin{figure} 
\centering
\includegraphics[width=1\columnwidth]{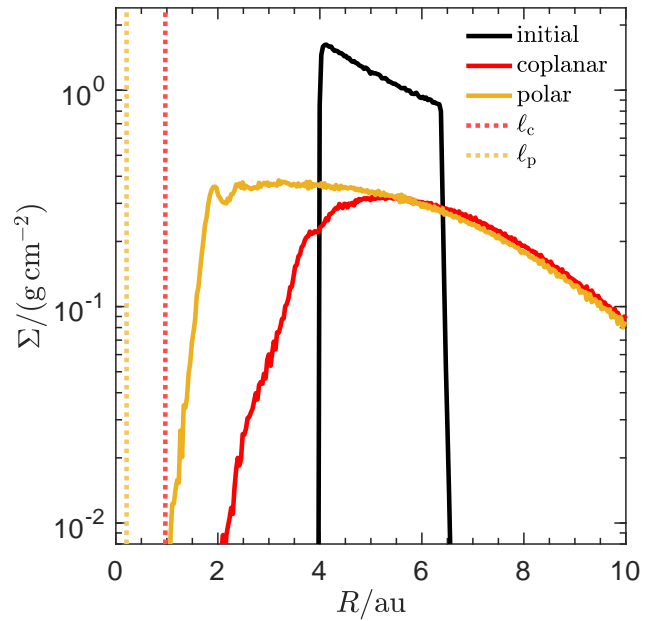}
\centering
\caption{The disc surface density profile at time $t = 1000\, \rm P_{orb}$ for a coplanar disc (red, run1 from Table ~\ref{table1}), and polar disc (yellow, run4). The black line gives the initial distribution.  The vertical dotted lines show the closest approach of the secondary companion in the plane of the disc, measured from the initial center-of-mass of the binary. The red dotted line shows this distance for the coplanar binary, $\ell_c$, and the yellow dotted line shows this distance for the polar binary, $\ell_p$.  The inner edge of the polar disc is closer in than the coplanar disc.}
\label{fig::sigma}
\end{figure}

\begin{figure*} 
\centering
\includegraphics[width=1\columnwidth]{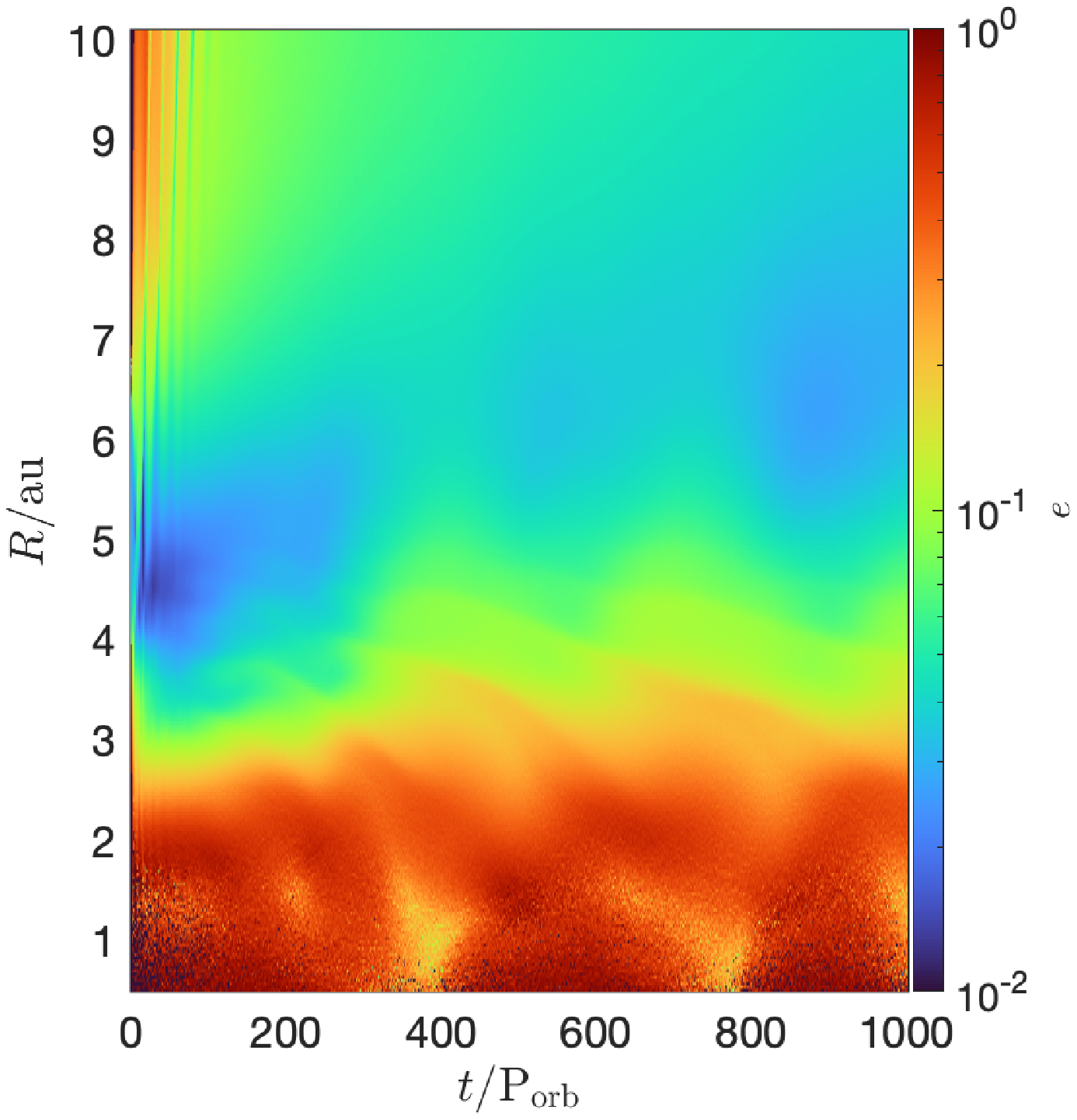}
\includegraphics[width=1\columnwidth]{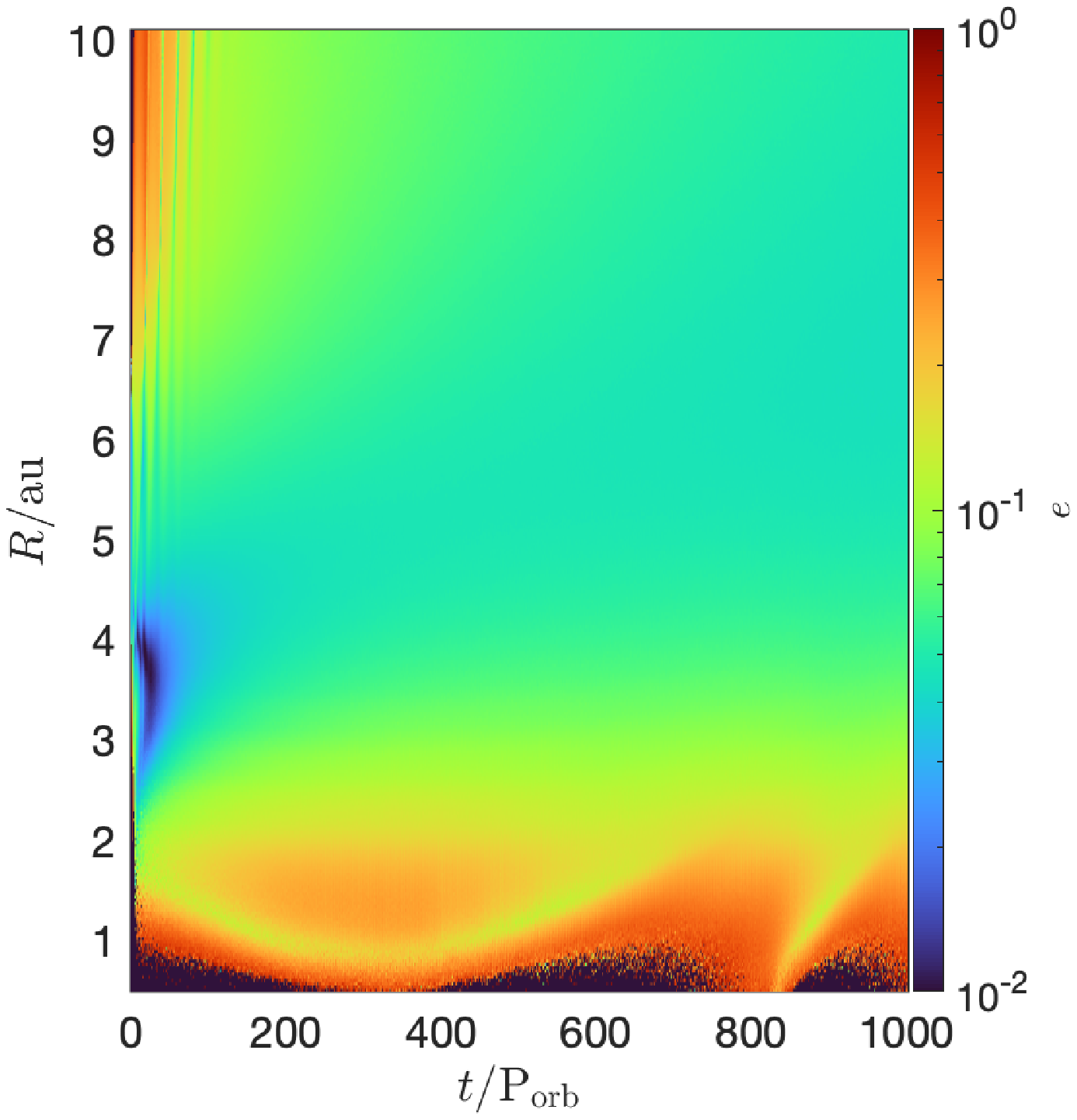}
\includegraphics[width=1\columnwidth]{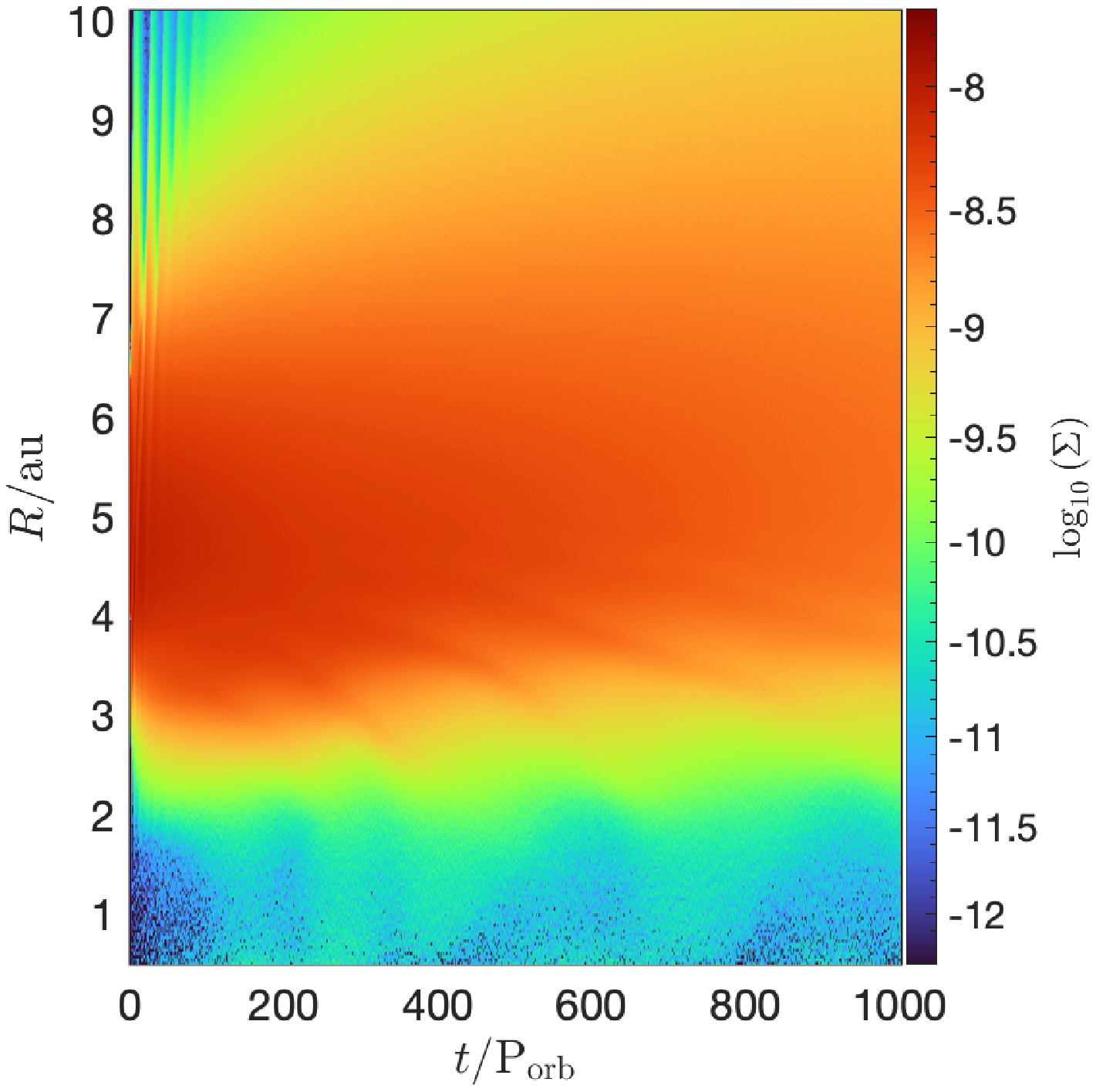}
\includegraphics[width=1\columnwidth]{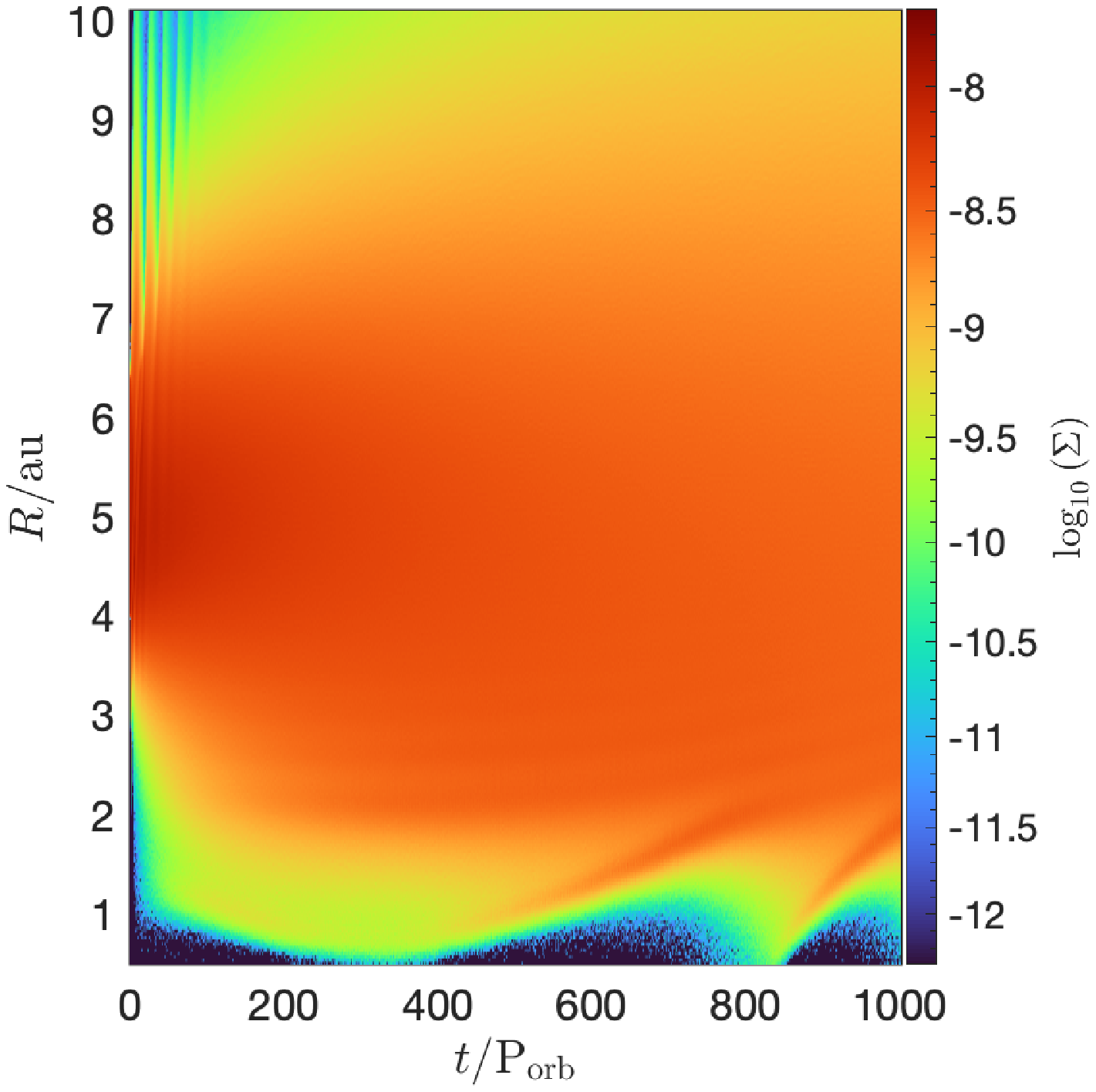}
\centering
\caption{The disc eccentricity evolution for a coplanar disc (upper left panel, run1 from Table ~\ref{table1}) and polar disc (upper right panel, run4). The $x$ and $y$ axes give the time in binary orbital periods, $P_{\rm orb}$, and disc radius, $r$, respectively. The upper colour bars denote the disc eccentricity value, $e$. Eccentric material arises within the coplanar disc that is not present in the polar disc. The bottom left and right panels show the disc surface density evolution for an initial disc tilt of coplanar and polar, respectively. The lower colour bars denote the disc surface density value, $\Sigma$. The disc that is initially polar is able to have stable material closer to the binary due to a weaker binary torque exerted onto the disc at a given radius.}
\label{fig::ecc_time}
\end{figure*}

\subsection{Circumbinary Disc Structure}
We investigate the  structure and evolution  of both an initially coplanar and polar circumbinary discs around an eccentric binary. Figure~\ref{fig::sigma} shows the surface density profile at $t = 1000\, \rm P_{orb}$ for the coplanar and polar models (runs 1 and 4 from Table ~\ref{table1}). The coplanar surface density profile is truncated at about $ 2- 3\, \rm au$, while the surface density profile for the polar disc extends inward to about $1\, \rm au$.   For a particular radius from the centre of mass of the binary, the tidal torque is stronger around a coplanar binary than a polar binary. As expected (see Section \ref{intro}), the polar disc's inner edge extends closer to the binary than the coplanar disc.   We also show the maximum orbital radius of the secondary component in the disc's plane for both coplanar and polar disc orientations. For the coplanar disc, this distance is $\ell_{\rm c} = 0.97\, \rm au$. For the polar disc, this distance is measured from the initial center-of-mass of the binary when the binary components are in the plane of the disc, which is  $\ell_{\rm p } = 0.21\, \rm au$.  %Given that the binary eccentricity in HD 98800BaBb is somewhat less than unity, the disc will have a reduced level tidal truncation by resonant torques in the polar case \citep{Franchini2019b}. 

%We show the evolution of the disc mass for both the coplanar and polar cases in Fig~\ref{fig::mass}. \SL{The the circumbinary disc is more rapidly depleted in mass in the coplanar case than the polar case. This likely because of the stronger tidal torque on the disc in the coplanar configuration.} After $1000\, \rm P_{\rm orb}$, the coplanar disc has lost $\sim 20$ per cent of its initial mass, while the initially polar disc has lost $\sim 8$ per cent. 

%\begin{figure} 
%\centering
%\includegraphics[width=1\columnwidth]{pattern_speed.eps}
%\centering
%\caption{The top panel shows the azimuthal position, $\theta$, of the high-density clump in the coplanar circumbinary disc (run1 from Table ~\ref{table1}) as a function of binary orbital periods, $P_{\rm orb}$. The bottom panel shows the patterns speed, $\omega$, of the clump in time.  }
%\label{fig::pattern_speed}
%\end{figure}

 \begin{figure*} 
\centering
\includegraphics[width=1\columnwidth]{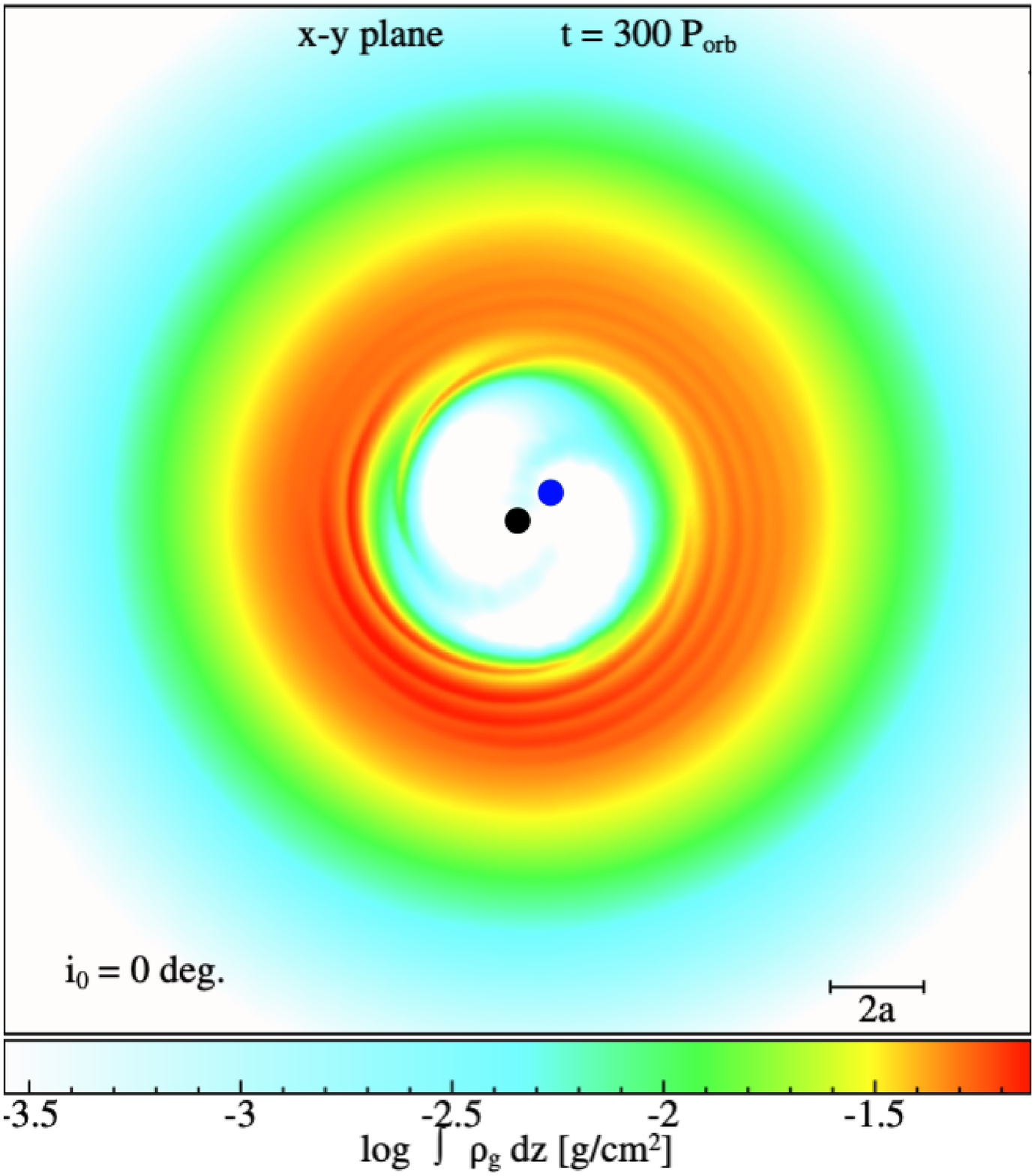}
\includegraphics[width=1\columnwidth]{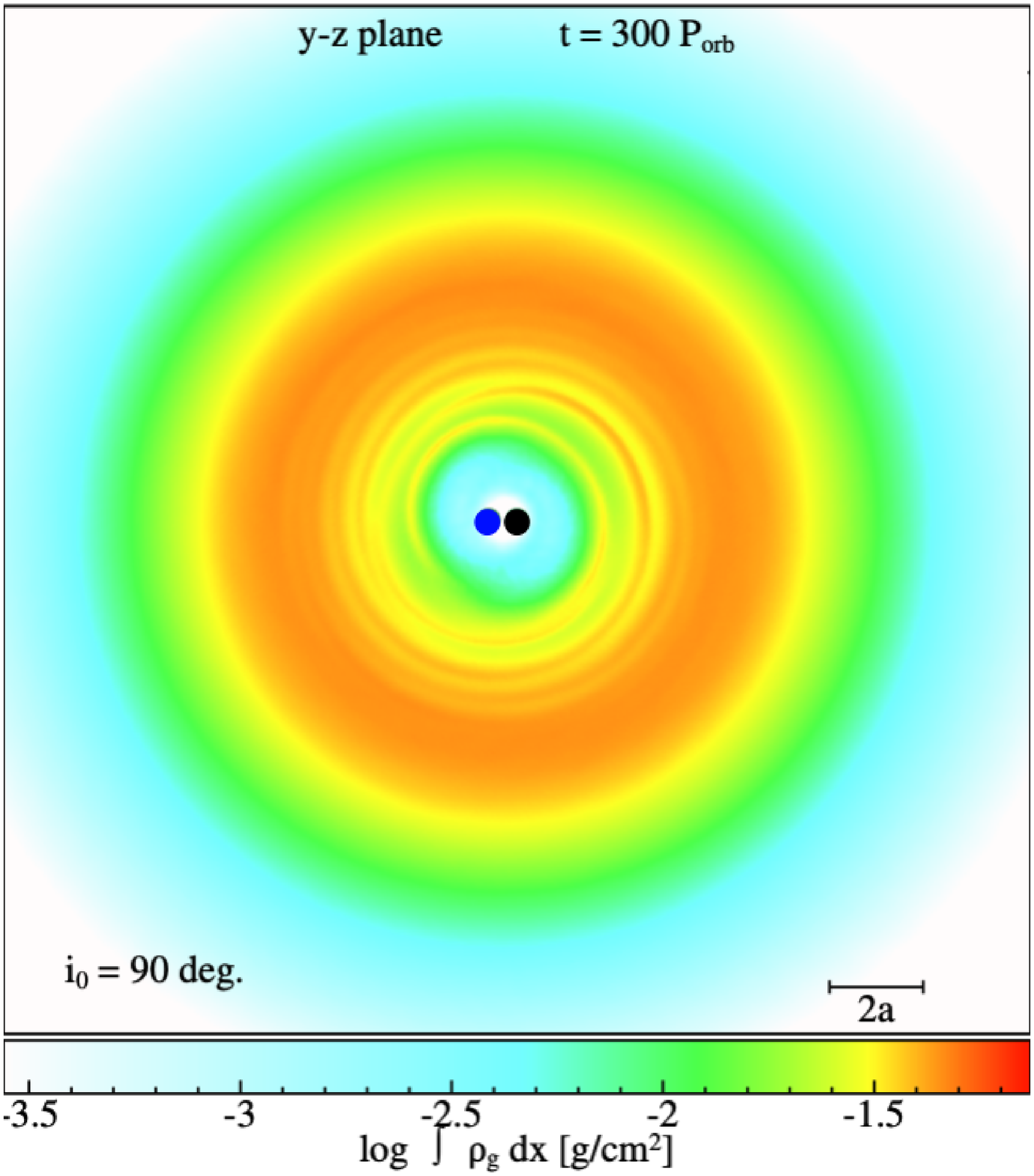}
\includegraphics[width=1\columnwidth]{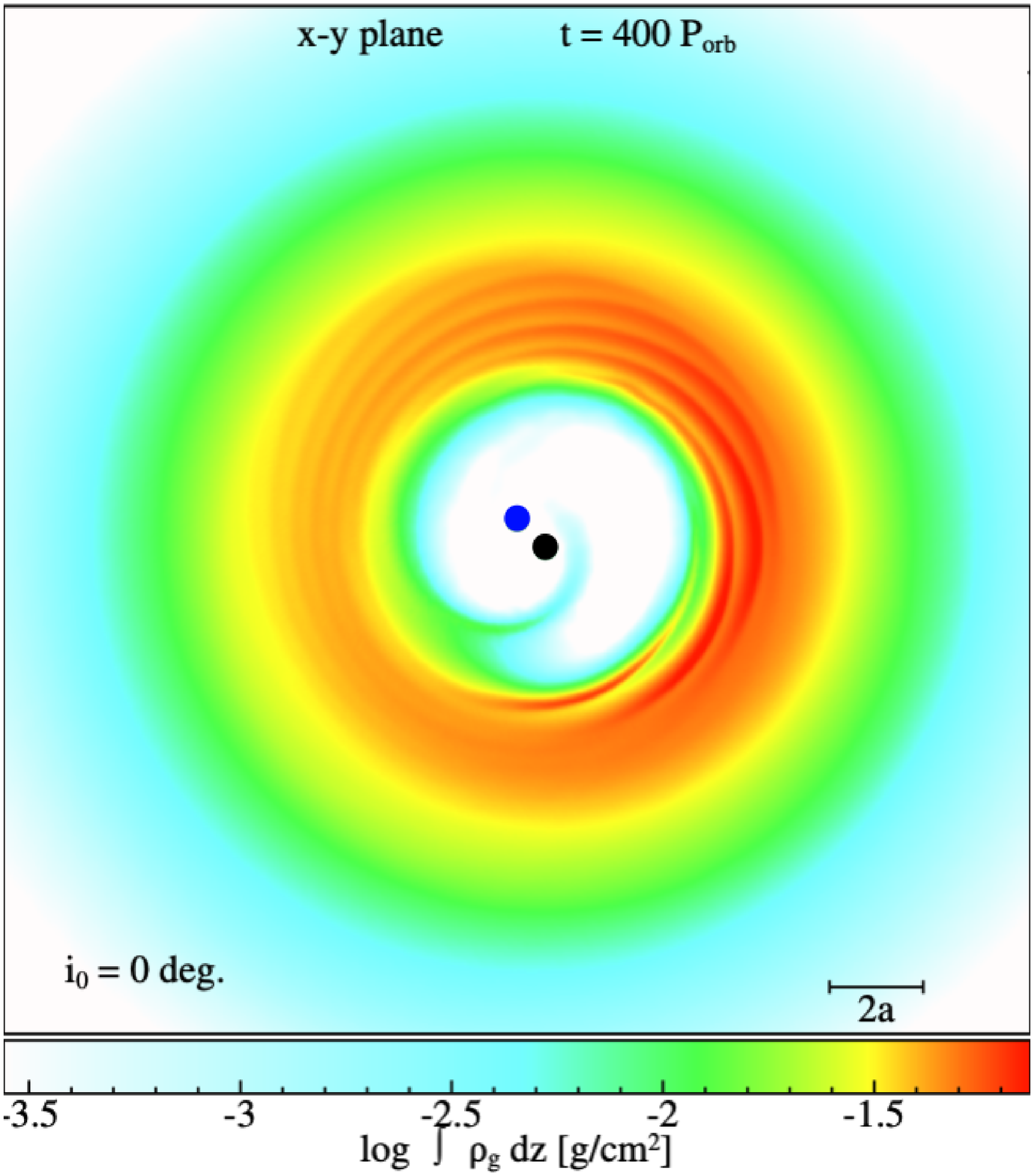}
\includegraphics[width=1\columnwidth]{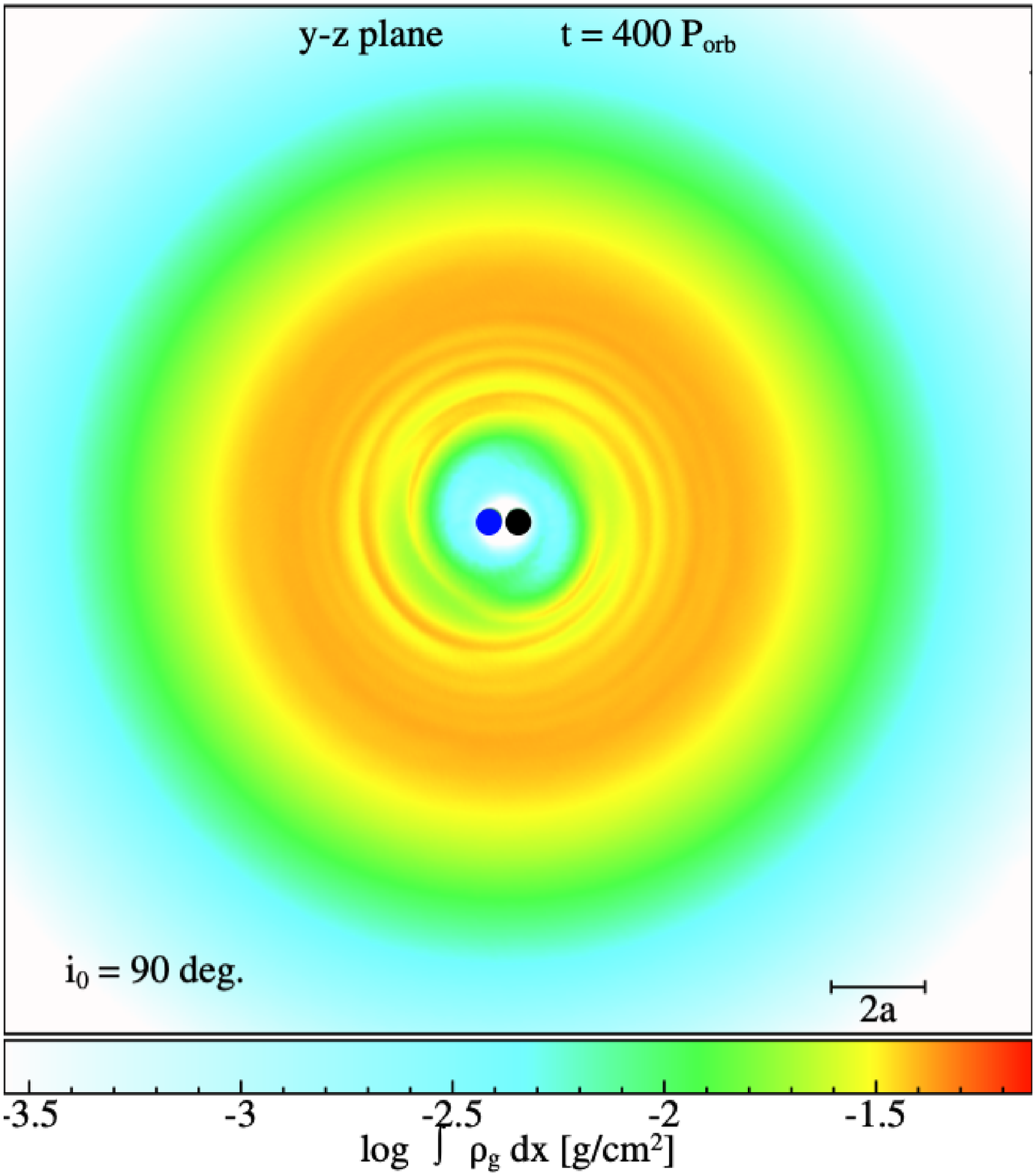}
\centering
\caption{The disc evolution for a hydrodynamical circumbinary disc at times $t = 300\, \rm P_{\rm orb}$ (top row) and $400\, \rm P_{\rm orb}$ (bottom row). The left-hand panels show a disc that is  coplanar ($i_0 = 0\degree$, run1 from Table ~\ref{table1}) and the right-hand panels show a disc that is  polar ($i_0 = 90\degree$, run4) to the binary orbital plane. The binary components are denoted by the circles with binary semi-major axis $a = 1\, \rm au$. The black circle represents the primary star and the blue denotes the secondary star.  The colour bar denotes the gas surface density. The coplanar disc is viewed looking down on the binary orbital plane, the $x$--$y$ plane. The polar disc is viewed in the $y$--$z$ plane. In both cases, the binary eccentricity vector is along the $x$--axis. At a time of $t = 300\, \rm P_{orb}$, a strong accretion stream can be seen flowing onto the secondary companion, while at $t = 400\, \rm P_{orb}$, the stream is flowing onto the primary. } 
\label{fig::splash}
\end{figure*}

We further investigate the disc eccentricity and surface density for the two disc configurations. The top row in Fig.~\ref{fig::ecc_time} shows the radial disc eccentricity evolution for the coplanar disc (run1 from Table~\ref{table1}) versus the polar disc (run4) and the bottom row shows the radial disc surface density as a function of time. There is heightened eccentricity growth around $2-3\, \rm au$ for the duration of the simulation for a coplanar disc. The patches of high eccentricity at $<2\, \rm au$ are in the central gap region (seen from the bottom-left panel) where the gas streams flow.  The polar-orientated disc shows no heightened eccentricity growth except at the inner disc edge. From the surface density renderings, we see more material closer to the binary in the polar disc than in the coplanar disc, as expected. Such eccentric disc behaviour around a binary at a similar binary eccentricity 
has been found previously, as discussed in Section~\ref{intro}. The cause of the disc eccentricity is not known. It could be due to tidal resonances with the binary \citep{Lubow1991}.
%{\bf Not sure about saying this: It might also be due  to gas streams that get ejected and collide with the inner edge of the circumbinary disc, as is known to occur in the circular orbit binary case \citep[e.g.,][]{MacFadyen2008, Shi2012,  Farris2014}.}

Fig.~\ref{fig::splash} shows the global disc structure for both  coplanar (left column) and polar (right colmun) orientations at $t = 300\, \rm P_{\rm orb}$ and $400\, \rm P_{\rm orb}$, respectively.  For the coplanar model, there are prominent spiral streams transporting material from the circumbinary disc through the binary cavity and eventually accreting onto the binary. At $t = 300\, \rm P_{\rm orb}$ the strong stream is flowing dominantly onto the secondary companion, while at $t = 400\, \rm P_{\rm orb}$ the stream is being dominantly accreted by the primary. For the polar model, no prominent streams are observed. Again, as expected, the cavity is larger in the coplanar case.  
%For the polar case, the binary torque is smaller at a given radius and this allows the inner edge of the disc to be stable closer to the binary \citep{Lubow2018,Franchini2019b}.

 Another difference between the two disc structures is a prominent asymmetric density enhancement that arises within the coplanar disc but not the polar disc. This density  feature is associated with the circumbinary disc becoming eccentric. The eccentricity is the largest at the disc's inner edge and then decreases in the outward radial direction (Fig.~\ref{fig::ecc_time}). This structure
 is a consequence of the trapped eccentric mode  structure within a circumbinary disc  \citep{Shi2012, Munoz2020b}.  We do not find the more detailed figure-eight like structures reported in 2D simulations by \cite{Mosta2019}. This may be because our simulations do not run as long and have lower spatial resolution
 in the gap region. On the other hand, these features
 may be less prominent in 3D that our simulations use.

\begin{figure} 
\includegraphics[width=\columnwidth]{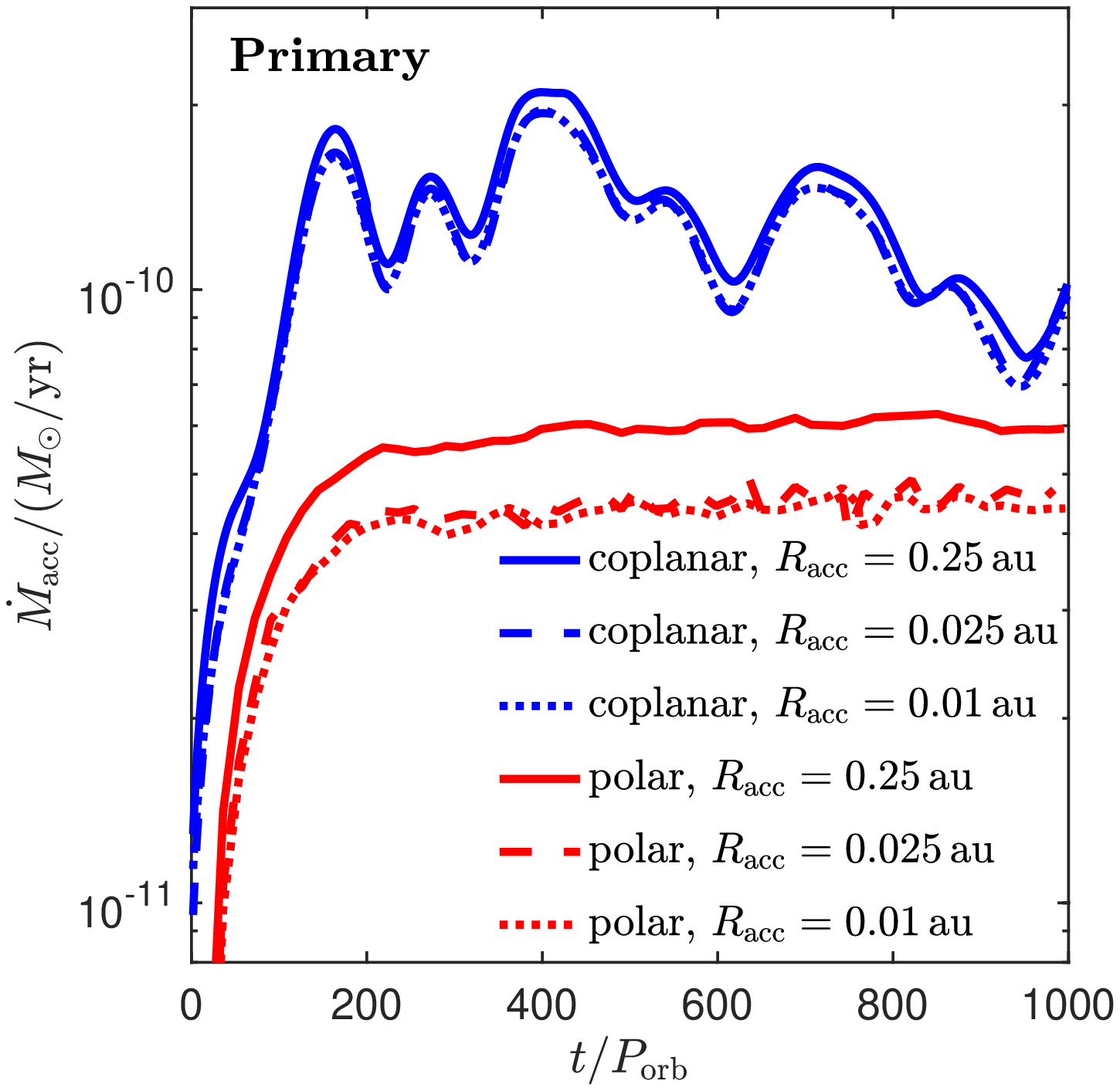}
\includegraphics[width=\columnwidth]{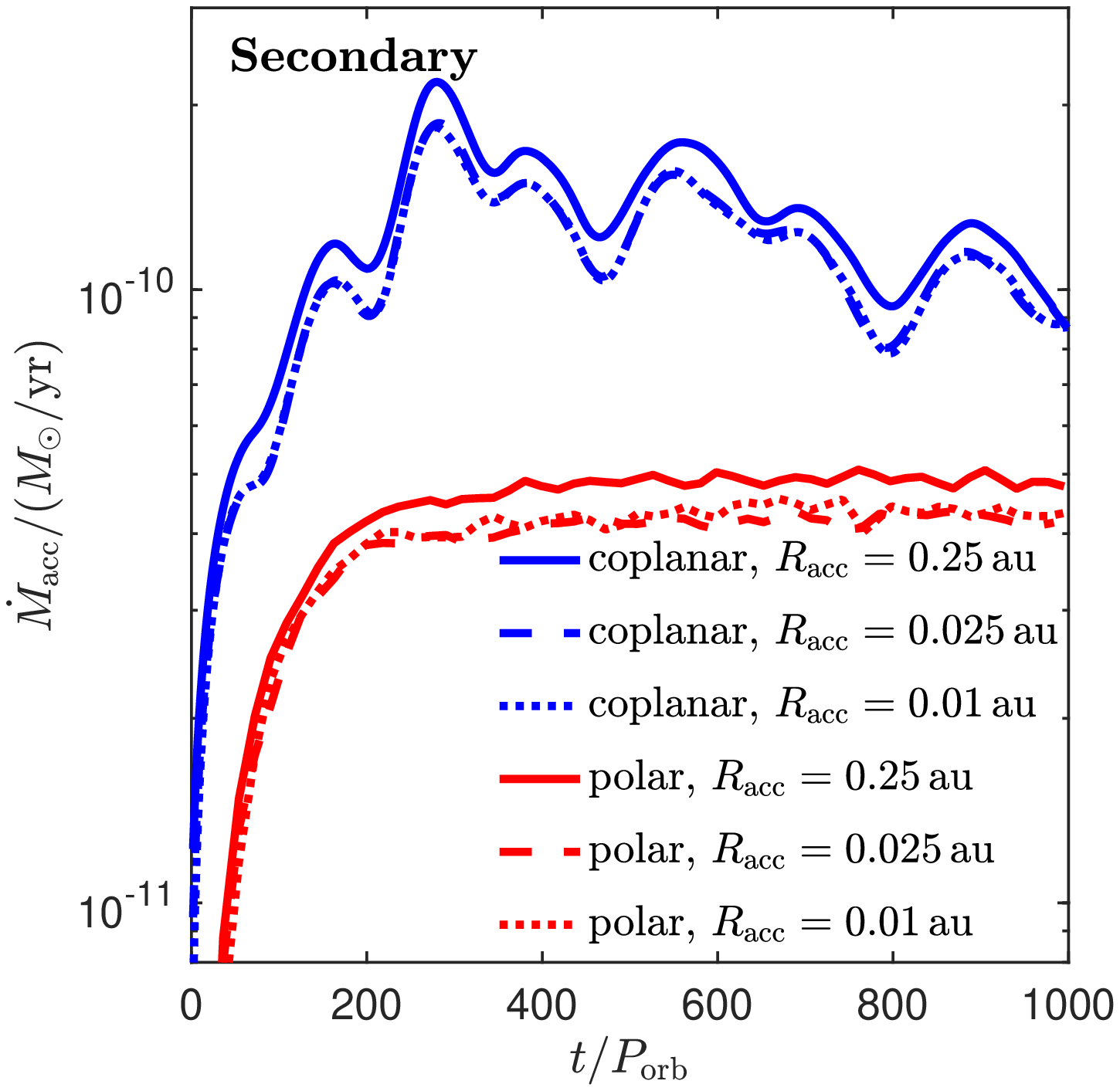}
\centering
\caption{The accretion rate, $\dot{M}_{\rm acc}$,  onto the primary star (top panel) and secondary star (bottom panel) as a function of  time in units of the binary orbital period for a our coplanar and a polar disc models.  The accretion rate is measured in solar mass per year, $M_{\odot}/\rm yr$. The disc orientation and accretion radius of each binary component is given by the legend.}
\label{fig::acc_rate_sims_pri}
\end{figure}

 \begin{figure} 
\centering
\includegraphics[width=\columnwidth]{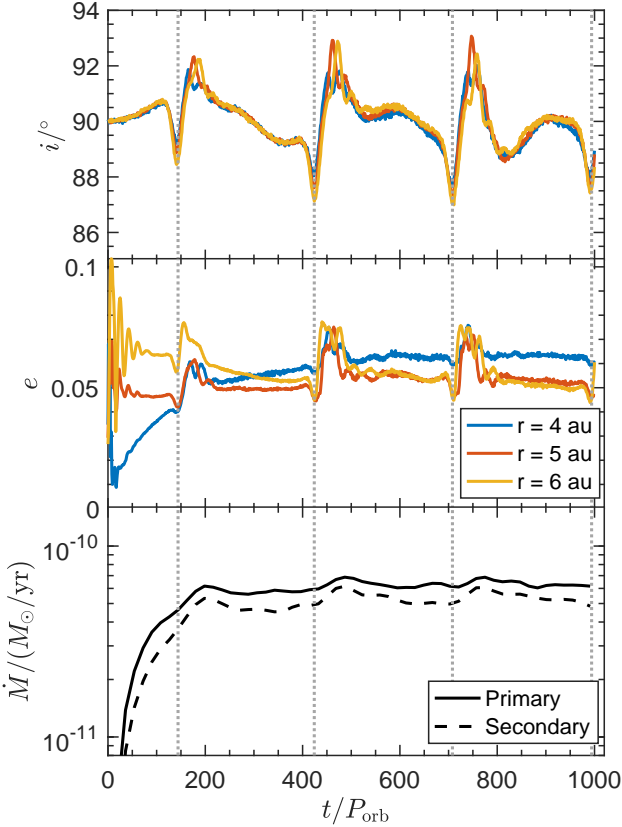}
\centering
\caption{ The evolution of the disc tilt, $i$, eccentricity, $e$, and accretion rate,  $\dot{M}_{\rm acc}$, as a function of  time in units of the binary orbital period in the presence of the HD 98800 A companion. For the disc tilt and eccentricity, we measure three different radii within the disc, $r=4,5,6\, \rm au$. The accretion rate onto the primary is shown by the solid line, while the secondary is shown by the dashed line.
 The vertical gray dotted lines represent the time of periastron passages for HD 98000 A.}
\label{fig::3star_data}
\end{figure}

\subsection{Accretion rates}

Next, we examine the accretion rates onto the primary and secondary stars for the coplanar and polar disc models. The upper panel of Fig.~\ref{fig::acc_rate_sims_pri} shows the accretion rate, $\dot{M}_{\rm acc}$, onto the primary star as a function of time.  For the both disc models, we test three different accretion radii $R_{\rm acc} = 0.25,\, 0.025,\, 0.01\, \rm au$.   In these simulations, the accretion rate initially increases over time due to the disc's inner parts viscously spreading inwards. The simulations with different accretion radii show similar rates, and so the accretion rate is approximately independent of the accretion radius. 
The accretion rate reaches a quasi-steady state in a time of about $150 - 200\,P_{\rm orb}$. No true steady state is possible because each disc loses mass over time. 
At a time of $1000 P_{\rm orb}$ the coplanar disc has lost about  $20\%$ of its initial mass, while the polar disc has lost about  $15\%$.  

The last column in Table~\ref{table1} shows the time-averaged accretion rates onto the primary and secondary binary components. For the coplanar models, the time-averaged accretion rate  onto the primary is similar to the rate onto the secondary when averaged over the oscillations.  For the polar models, the accretion rates onto the binary with the largest accretion radius are less than a factor of 1.4 greater than the smallest accretion radius. Again showing that the accretion rate does not depend strongly on the accretion radii of the stars.

In the coplanar case, the accretion rates onto the primary and secondary oscillate after a time of about $150 \,P_{\rm orb}$, while the global accretion rate decreases over time.
The oscillations in accretion rate are due to the apsidal precession of the circumbinary disc after it becomes eccentric (see Fig.~\ref{fig::ecc_time}) and occurs on a timescale of $\sim 150\, P_{\rm orb}$.
The accretion rates onto the primary and secondary are anti-phased. That is, the accretion rate onto the primary is at a maximum when the accretion rate onto the secondary is at a minimum. 
Such oscillations, with a similar oscillation timescale, were found in previous work \citep[e.g.,][]{Munoz2016,Munoz2019}.

In the polar case, no significant accretion oscillations occur and the disc retains its nearly circular form.
The accretion rate in the polar case does not decrease in time over the simulated time range, but must eventually decrease in time as the circumbinary
disc mass becomes depleted.  The accretion rates onto the primary and secondary are nearly equal.
At a time of $1000\, \rm P_{orb}$, the binary accretion rate in the coplanar case is about a factor of two larger than in the polar case. 

 The gap sizes in the coplanar and polar cases are quite different (see Fig.~\ref{fig::splash}), although
the accretion rates are somewhat similar and appear to be approaching each other. This result may suggest that these cases are evolving towards  steady state accretion through the circumbinary disc and into the gap (averaged over short timescale accretion modulations), as has been suggested in recent studies \citep[e.g.,][]{Munoz2019, Dittmann2022}. In such cases, the accretion rate onto the binary would then be independent of the binary properties and resulting gap size.

 \begin{figure*} 
\centering
\includegraphics[width=1\columnwidth]{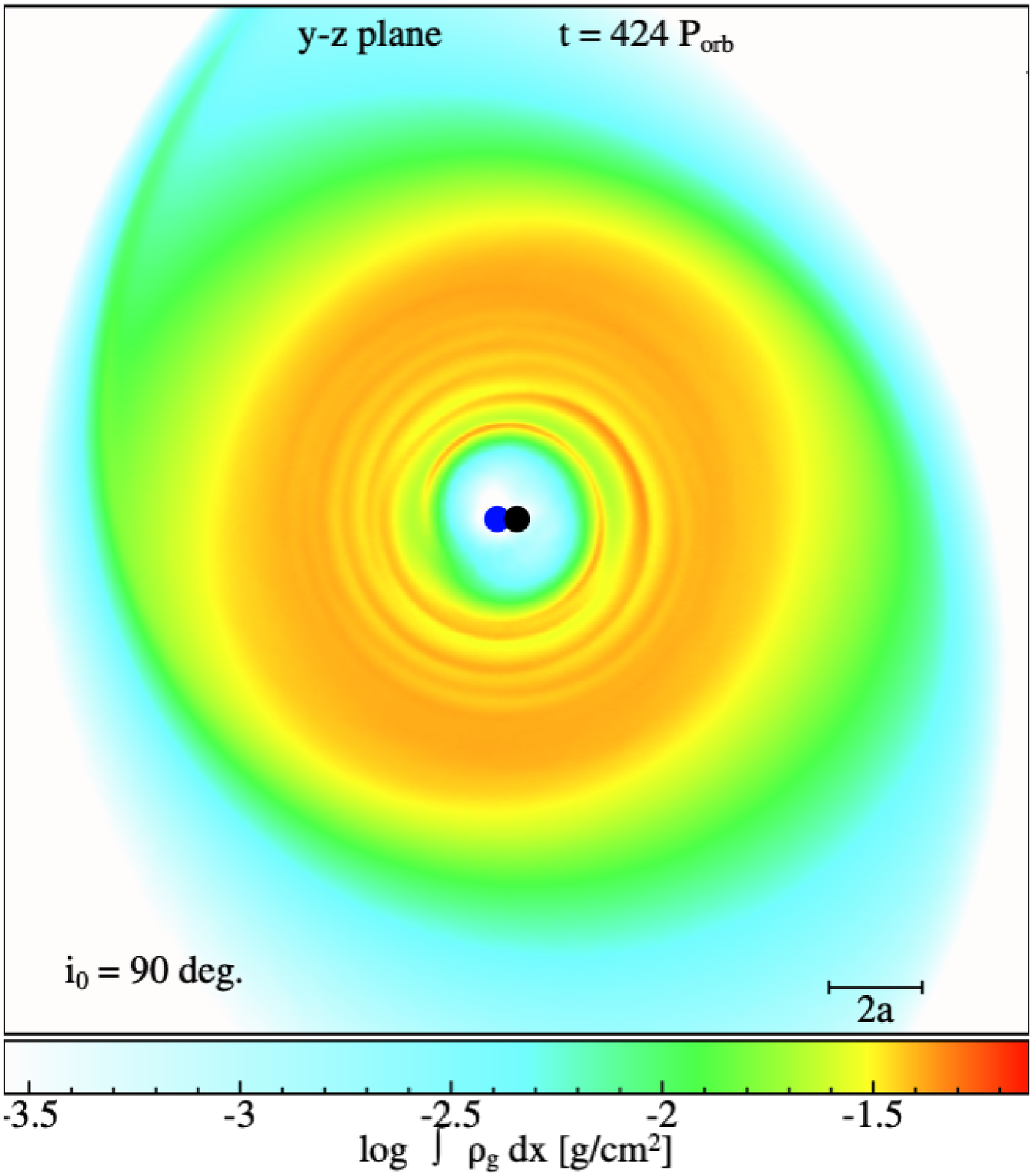}
\includegraphics[width=1\columnwidth]{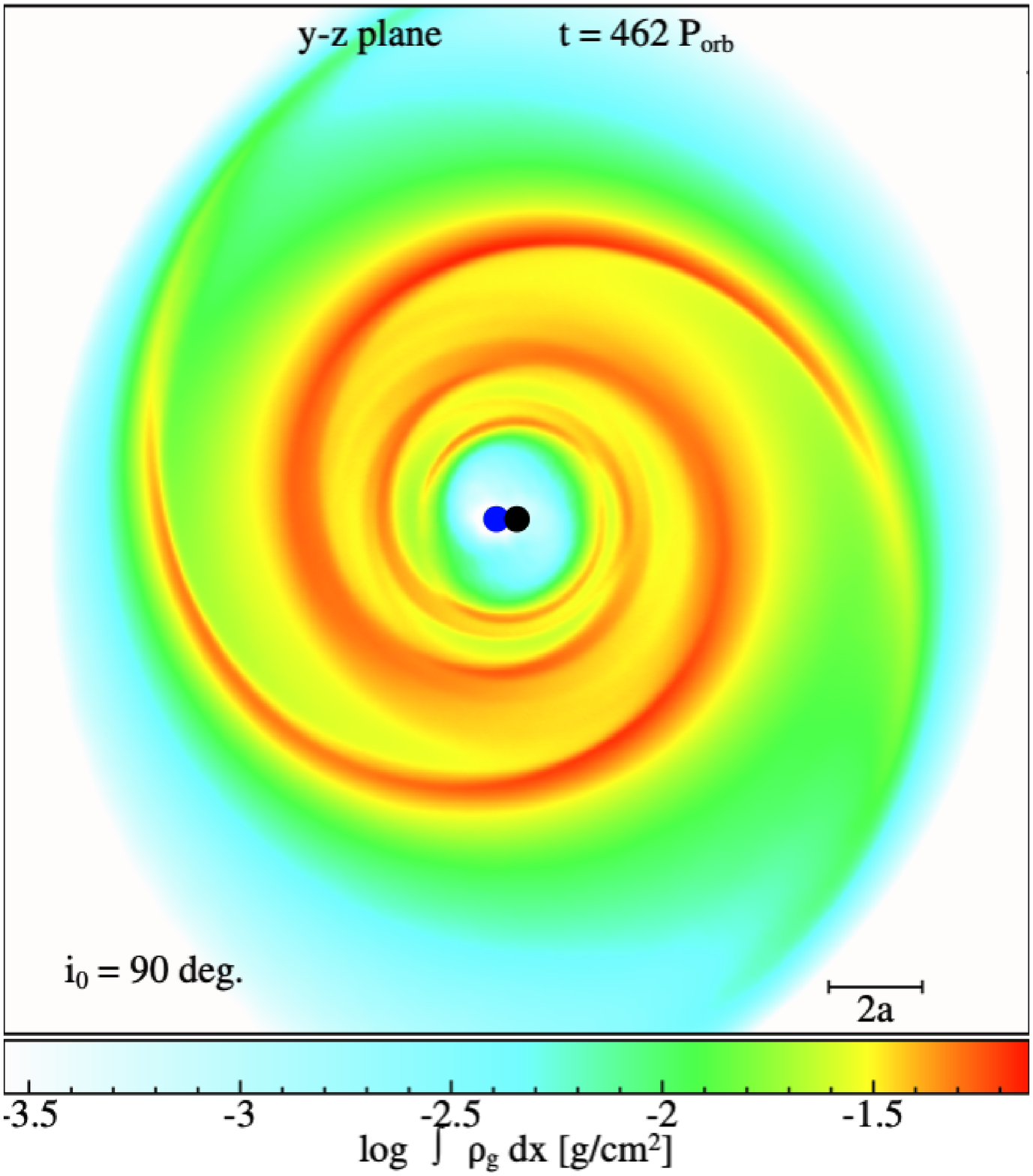}
\centering
\caption{ The disc evolution for a polar circumbinary disc around HD 98800 BaBb perturbed by HD 98800 A (run7) at times $t = 424\, \rm P_{\rm orb}$ (periastron passage of HD 98800 A, left panel) and $462\, \rm P_{\rm orb}$ (time shortly after periastron, right panel). The black circle represents the primary star and the blue denotes the secondary star.  The colour bar denotes the gas surface density. The circumbinary disc is viewed in the $y$--$z$ plane, with the binary eccentricity vector is along the $x$--axis. At a time of $t = 462\, \rm P_{orb}$, prominent spiral arms are formed by HD 98800 A.}
\label{fig::3star_splash}
\end{figure*}

\section{Effect of the outer companion HD 98800 A}
\label{sec::AB}
 HD 98800 B is part of a quadruple system (as described in Section~\ref{intro}). We run an additional simulation including the HD 98800~A outer companion in order to examine the effect that it has on the accretion rate onto the B binary. We simulate a polar circumbinary disc around the HD 98800 B binary using the same disc and binary parameters as in Section~\ref{sec::Hydro_Sims}. We model the HD 98800 A binary as a single star by combining the masses of the individual Aa and Ab stars. We set the semi-major axis and eccentricity to $a_{\rm A} = 54\, \rm au$ and $e_{\rm A} = 0.52$, respectively \citep{Kennedy2019}. We calculate the tilt $i_{\rm A}$, argument of the pericentre $\omega_{\rm A}$, and longitude of the ascending node $\Omega_{\rm A}$ of HD 98800 A in the AB frame based on the binary properties in the sky plane from \cite{Kennedy2019}. We find that  $i_{\rm A} = 34^\circ$, $\omega_{\rm A} = 137^\circ$, and $\Omega_{\rm A}= 125^\circ$. We use an accretion radius of $0.25\, \rm au$ for the B binary components, and use $1.85\, \rm au$ for the A companion. The outer companion does not significantly affect the disc dynamics for this system because the inner binary torque on the disc dominates the outer companion torque \citep{Verrier2009,Martin2022}.

Figure~\ref{fig::3star_data} shows the disc tilt and eccentricity, as well as the accretion rate onto the B binary components, as a function of time in units of the B binary orbital period. We show the times of the periastron passage of HD 98800A by the vertical dotted lines. During each periastron passage, the disc tilt decreases by a few degrees and then increases soon after the periastron of the A companion. This may be why we observe the disc tilt to be $\sim87^\circ$ rather than exactly polar. Shortly after the periastron passage, the companion excites eccentricity growth in the polar circumbinary disc, temporarily enhancing the accretion rate. The accretion rate is on the order of $10^{-11} M_{\odot}/ \rm yr$, which is comparable to the simulation without the outer companion. However, the accretion rate modulates in time, increasing by $\sim 20$ per cent after each periastron encounter of the companion. The primary accretion is higher than the secondary accretion, which is consistent with our previous simulations.

We analyze the global disc structure during the time of the closest approach of the A companion. Figure~\ref{fig::3star_splash} shows the disc structure at $424\, \rm P_{orb}$ (periastron, left panel) and $462\, \rm P_{orb}$ (shortly after periastron passage, right panel). There are no observed changes in the overall disc structure when the companion is at periastron. However, shortly after periastron, the companion excites prominent spiral arms in the disc, which temporarily enhances the accretion rate (see Fig.~\ref{fig::3star_data}).

\section{Accretion onto  HD 98800 B\lowercase{a}B\lowercase{b}}
\label{sec::Mass_accretion}
%The long-term evolution of the binary orbit is dictated by the interaction between the two central bodies and the surrounding gas via accretion and gravitational torques. The tidal torque of the binary exerted on the disc can facilitate, and perhaps limit, the process of accretion. Usually, the accretion occurs in short, high-speed streams originating along the outside of the cavity and spiraling inward to the central binary. The mass accretion rate onto a star, $\dot{M}_{\rm acc}$, is instrumental for the study of accretion disc evolution.
 %This parameter is important for disc evolution models, which coincide with pre-main sequence stellar evolution and planet formation \citep{Hartmann1998}. 
 Observationally, $\dot{M}_{\rm acc}$ can be estimated by measuring the flux of continuum and line emission that is due to the shock of  infalling gas from a disc onto the central star along the stellar magnetic field lines \cite[e.g.,][]{Calvet1998}. The release of the accretion energy from the shock in the form of line emission is also known as the accretion luminosity $L_{\rm acc}$ \citep{Gullbring1998,Hartmann1998}.  The accretion luminosity $L_{\rm acc}$ is often computed from empirical relations between line luminosity, $L_{\rm line}$, and $L_{\rm acc}$ \cite[e.g.,][]{Natta2006,Biazzo2012,Fang2013,Antoniucci2014,Manara2015}.

From observations, the emission line fluxes $F_{\rm line}$ can be calculated from the calibrated spectrum. From this, the line luminosities are given by
\begin{equation}
    L_{\rm line} = 4\pi D^2 F_{\rm line}
\end{equation}
\citep{Rigliaco2012}, where $D$ is the distance to the source. The accretion luminosities can be determined by using the measured line fluxes according to empirical relations between the observed line luminosities and the accretion luminosity or mass accretion rates \citep{Gullbring1998,Natta2004,Mohanty2005,Herczeg2008,Fang2009}. The empirical relationships were calibrated on small samples of well-observed T Tauri stars and brown dwarfs \citep{Gullbring1998,Muzerolle2003,Calvet2004}. The distance to HD 98800 BaBb is $D = 47\, \rm pc$ \citep{VanLeeuwen2007}. The brightest component has a $\rm H\alpha$ flux of $6.5\times 10^{-16 } \, L_{\odot}$ \citep{Zurlo2020}.  Given this flux value, the line luminosity is $L_{\rm line} = 4.6 \times 10^{-8} \,  L_{\odot}$ \citep{Zurlo2020}.

The empirical relationship for $L_{\rm acc}$ based on  $L_{\rm line}$ is given by
\begin{equation}
    \log(L_{\rm acc}/L_{\odot}) = b + a \times \log(L_{\rm line}/L_{\odot})
    \label{eq::L_acc}
\end{equation}
\citep{Rigliaco2012}, where $a$ and $b$ are the line luminosity vs. accretion luminosity relationships for selected accretion indicators \citep[e.g.][]{Fang2009}. The accretion indicator that we are interested in is the $\rm H\alpha$ line. Given that $a = 1.49$ and $b = 2.99$ for  the $\rm H\alpha$ line and $L_{\rm line} = 4.6 \times 10^{-8} \,  L_{\odot}$, the accretion luminosity for the brightest component in HD 98800 BaBb is $L_{\rm acc} = 1.14 \times 10^{-8}\, \rm L_{\odot} $. This value agrees with the findings of the accretion luminosity from \cite{Zurlo2020}. Once $L_{\rm acc}$ is estimated, it can be converted into a mass accretion rate, $\dot{M}_{\rm acc}$, given by
\begin{equation}
    \dot{M}_{\rm acc} = \bigg(1-\frac{R_{*}}{R_{\rm in}} \bigg)^{-1} \frac{L_{\rm acc} R_{*}}{G M_{*}}
    %\sim 1.25\frac{L_{\rm acc} R_{*}}{G M_{*}},
    \label{eq::M_acc}
\end{equation}
\citep{Rigliaco2012}, where $G$ is the universal gravitational constant,  $R_\star$ is the radius of the star, and $R_{\rm in}$ is the inner truncation radius for the disc. The binary components will accrete material from a circumstellar discs that is continuously fed by a circumbinary disc. The factor $\big(1-\frac{R_{*}}{R_{\rm in}} \big)^{-1} \sim 1.25$ is estimated using $R_{\rm in} \sim 5\, R_{*}$, which assumes that the accretion gas falls onto the star from the truncation radius of the circumstellar disc \cite[e.g.,][]{Gullbring1998}.  
Using $L_{\rm acc} = 1.14 \times 10^{-8}\, \rm L_{\odot}$, $R = 1.09\, \rm R_{\odot}$, and $M_{*} = 0.699\, \rm M_{\odot}$ \citep{Boden2005}, we estimate the mass accretion to be $6.8\times 10^{-16}\, \rm M_{\odot}/\rm yr$.

For comparison, the accretion rates for single T Tauri stars are typically in the range of $\sim 10^{-10} - 10^{-7}\, \rm M_{\odot}/yr$ \citep{Valenti1993,Gullbring1998,Calvet2004,Ingleby2013}. Observations reveal accretion rates onto binary star systems are comparable to that of accretion rates onto single T Tauri stars \citep{White2001}. 
Binary components and single stars of similar mass have similar mass accretion rates. However, it has been shown that more massive stars generally have larger accretion rates than less massive stars \citep{White2001}. %\cite{Alves2019} estimated the mass accretion rate from the circumbinary disc onto the circumstellar discs in [BHB2007] 11B to be $\sim 1.1 \times 10^{-5}\, \rm M_{\odot}/yr$. While the orientation of the [BHB2007] 11B circumbinary disc is uncertain, the relatively high accretion rate and formation of circumstellar discs suggest that this disc is in a non-polar state.  
In HD 98800 BaBb, only the primary star (the slightly more massive binary component) has detectable $\rm H\alpha $ flux. 

The mass accretion rate onto the primary from the simulation of a polar disc with the HD 98800 A companion (run7) is $\sim 6\times 10^{-11}\, M_{\odot}/ \rm yr$. Given that $\dot{M} \propto \alpha (H/R)^2$, a reduction in $\alpha$ and $H/R$ is needed to obtain the observed accretion rate of  $\sim 7\times 10^{-16}\, \rm M_{\odot}/\rm yr$ for the HD 98800 BaBb polar circumbinary disc. Therefore, we require $\alpha < 10^{-5}$ and $H/R < 0.05$. 
Such low values would be inconsistent with the tilt evolution timescale to a polar state. Linear models in \cite{Lubow2018}
suggest a tilt evolution timescale of $\tau \sim 5 \times 10^4 (H/R)^2/\alpha \,  \rm{yr}$ that could exceed the age of the system that is estimated as $\sim 10^7 \, \rm{yr}$ \citep{Kennedy2009,Ronco2021} (nonlinear effects could reduce this timescale).
Our models also predict that the accretion rate onto the secondary is about the same as the accretion rate onto the primary. Therefore it is unclear why emission from only the primary is detected.

\section{Conclusions}
\label{sec::conclusions}
 We have simulated the accretion of coplanar and polar circumbinary gas onto a highly eccentric binary with a mass ratio close to unity. 
 For the same disc properties, except inclination, both the coplanar disc and polar disc accrete onto the primary and secondary stars at similar rates when averaged over many binary
 orbital periods. 
 The  polar disc has significantly different properties from the coplanar disc. Unlike the coplanar disc, the eccentricity of the polar disc does not grow.
  For a coplanar disc, the accretion rates  onto the binary components undergo oscillations  so that the dominant accretion alternates between the components as a 
 consequence of the precession of the eccentric disc, as was previously found \citep{Munoz2016, Munoz2019}. However, for a polar disc, no such oscillations are found as a consequence of its persistent nearly circular form.

 We applied our findings to the eccentric binary system, HD 98800 BaBb, which contains a polar circumbinary gas disc.  We include the outer companion HD 98800 A to determine its effects on the accretion rate onto the B binary during the periastron passage of the A companion. We found that the eccentric orbit of HD 98800 A increases the accretion rate onto HD 98800 B by $\sim 20$ per cent after each periastron passage.  Using the observations from \cite{Zurlo2020}, we estimated the primary mass accretion rate to be $\dot{M}_{\rm acc} = 6.8 \times 10^{-16}\, \rm M_{\odot}/\rm yr$. 
This accretion rate is about $6$ orders of magnitude lower than observations of accretion rates in  T Tauri stars.  Our hydrodynamical simulations are unable to explain such a low accretion rate unless the $\alpha$ parameter is very small, $\alpha < 10^{-5}$. %, which may be reasonable for protoplanetary discs.} 
 Furthermore, we predict similar accretion rates onto both binary components, while observational signatures
of accretion were found for only the primary.  Further observational checks on the accretion rate 
would help to clarify these differences.

%We have simulated the accretion of misaligned circumbinary material on to an eccentric binary. Circumbinary discs that are either in a coplanar or nearly coplanar configuration result in an accretion rate that undergoes alternating preferential accretion. This phenomenon is produced from the disc becoming eccentric and forming a high-density clump.  On the other hand, a disc in a polar state, or aligning to polar, does not show this accretion variability. However, primary accretion is always higher than secondary accretion when the disc is polar aligned.  {\bf The total accretion rate on to a polar binary is only slightly lower than that onto a coplanar binary.}

\section*{Acknowledgements} 

 We thank the referee for helpful suggestions that positively impacted this work. We thank Daniel Price for providing the phantom code for SPH simulations and acknowledge the use of splash \citep{Price2007} for the rendering of the figures. Computer support was provided by UNLV's National Supercomputing Center. We acknowledge support from NASA through grants 80NSSC21K0395 and 80NSSC19K0443.This research was supported in part by the National Science Foundation under Grant No. NSF PHY-1748958.

\section*{Data Availability}

The data supporting the plots within this article are available on reasonable request to the corresponding author. A public version of the {\sc phantom} and {\sc splash} codes are available at \url{https://github.com/danieljprice/phantom} and \url{http://users.monash.edu.au/~dprice/splash/download.html}, respectively.

%%%%%%%%%%%%%%%%%%%%%%%%%%%%%%%%%%%%%%%%%%%%%%%%%%



\bibliographystyle{mnras}
\bibliography{ref} % if your bibtex file is called example.bib

\bsp	% typesetting comment
\label{lastpage}
\end{document}